\documentstyle[pre,aps,preprint]{revtex}

\font\helv=cmssbx10
\begin{document}
\draft
\title{Semiclassical Theory for Many-Body Fermionic Systems}
\author{ Pierre {\rm Gaspard} and Sudhir R. {\rm Jain} \\
 Facult\'e des Sciences and \\
Center for Nonlinear Phenomena and Complex Systems, \\
 Universit\'e Libre de Bruxelles,
Campus Plaine C.P. 231, \\ Boulevard du Triomphe, 
1050 Bruxelles, Belgium }
\maketitle

\begin{abstract}
We present a treatment of many-body Fermionic systems that facilitates an
expression of the well-known quantities in a series expansion in $\hbar $.  The
ensuing semiclassical result contains to a leading order of the response
function the classical time correlation function of the observable followed by
the Weyl-Wigner series, on top of these terms are the periodic-orbit correction
terms. The treatment given here starts from linear response assumption of the
many-body theory and in its connection with semiclassical theory, it makes no
assumption of the integrability of classical dynamics underlying the one-body
quantal system. Applications of the framework are also discussed.   
 
\end{abstract}
\newpage

\noindent
{\bf ~1. Introduction}

Semiclassical framework for single-particle systems is presently in an advanced
stage \cite{gab,bj}. For both the integrable and chaotic dynamics, we now
understand the semiclassical quantization. However, for intermediate \cite{jl}
and mixed dynamical scenario, we still lack a convincing theory. Another
situation of a great practical relevance arises in the systems comprising  of
many bodies. That the spectral fluctuation characteristics of these systems are
modelled in much the same way as for chaotic systems with lesser degrees of
freedom has been shown quite recently \cite{gv}. It is very important to note
that complex systems do possess features, mainly associated with the generic
nature of the thermodynamic limit, which are completely absent from the systems
with fewer freedoms, friction being one of the examples. There are other systems
which one would like to understand semiclassically. Some notable examples are of
quantum dots \cite{expt}, metallic clusters \cite{brack} and their optical
properties  \cite{dmb}; also, nuclear physics at high spins \cite{jja} presents
us with opportunity to test various semiclassical ideas in many-body theory.
There has been an attempt based on the semiclassical limit of the time-dependent
Hartree-Fock equation where one can establish a connection \cite{bdd} between
the strength function given by the Vlasov equation and the corresponding quantum
function in the limit of large quantum numbers. Analogous to the many-body
treatment based on random-phase approximation (RPA), the linearized Vlasov
equation gives an integral equation for the particle-hole propagator in terms of
classical propagator in the static mean field and two-particle Coulomb
interaction. Thus one needs to evaluate the classical propagator in the mean
field and then solve the integral equation.

There is a lot of evidence that periodic orbits in the semiclassical trace
formulae describe the shell effects in metallic clusters \cite{nhm}. Of course,
these have been preceded by nuclear physicists \cite{strut} by more than a
decade. In fact, the periodic-orbit theory \cite{mcg}, which, in principle,
should be rendered useless for practical purposes due to a huge number of
technical problems associated with finding periodic orbits and summing a
conditionally convergent series, gets encouragement from observations that only
a few periodic orbits are enough in many situations \cite{mcg}.

We present here a semiclassical treatment of the response function. Response
function is essentially the imaginary part of generalized dynamical
susceptibility which, in turn, is related to the correlation commutator. In the
section 2, we present the general discussion from the linear response theory to
arrive at a quantity for which semiclassical expression can be written.  The key
point of this section is to show that  the most important quantity is a time
correlation function. In the section 3, we present semiclassical treatment of
the system perturbed by an external influence. This enables us to write the
relevant results of sections 2 and 3 in a semiclassical expansion. We will see
that the leading term is indeed the two-time correlation function averaged over
the phase space, followed by the Weyl-Wigner series which has the periodic-orbit
corrections. On our way, we take account of the Fermionic nature of the
particles. \\  

\noindent {\bf ~2. Response Function and the Time Correlation Function}

We present here a discussion on the response of a system in the presence of an
external field. We work with susceptibility \cite{balescu} and express it in
terms of trace over single-particle states of two-time correlation function of
the observable. This quantity is also known as the polarization propagator
\cite{fw}.

If a system described by the Hamiltonian $\hat{H}$ is externally  disturbed by
an  field $F^{ext}(t)$, then the total Hamiltonian is  
\begin{equation}
\hat{H}_T = \hat{H} - \hat{Q}F^{ext}(t)
\end{equation}
where $\hat{Q}$ is an observable, an example could be magnetization in the
context of spin systems, or, an electric dipole operator in an example involving
photoabsorption, and so on. The response function can be written as the
imaginary part of the dynamical susceptibility, 
\begin{eqnarray} 
\chi"(t,t') &=&
(2\hbar )^{-1}\langle[\hat{Q}(t),\hat{Q}(t')]\rangle \nonumber \\ &=& \int
~\frac{d\omega }{2\pi }\ e^{-i\omega (t-t')} ~{\tilde \chi}"(\omega ).
\end{eqnarray} 
The angular brackets denote the expectation value and the square
brackets denote the commutator. Setting an initial time to $0$ and the final
time to $t$, we can write 
\begin{equation} 
\chi"(t) = (2\hbar )^{-1}\left[
\langle\hat{Q}(t)\hat{Q}(0)\rangle - \langle\hat{Q}(0)\hat{Q}(t)\rangle \right]
\end{equation} 
where
\begin{equation}
\hat{Q}(t) = \exp (i\hat{H}t/\hbar )\ \hat{Q}\ \exp (-i\hat{H}t/\hbar )
\end{equation}
and where $\langle\cdot\rangle$ denotes the average over the initial state of the
system which is, for instance, the thermal state wherein 
\begin{equation}
\langle\cdot\rangle = {1 \over{Z(\beta )}} \ {\rm tr}~e^{-\beta \hat{H}}(\cdot)
\end{equation}
with 
\begin{equation}
Z(\beta ) = {\rm tr}~\exp (-\beta \hat{H}),
\end{equation}
if the system was in contact with a thermal reservoir at temperature $T=1/\beta
k_B$ in the period preceding the interaction with the external field,
$F^{ext}(t)$. If the system was in its pure ground state $|\Phi _0\rangle$ then the
average is over this many-body eigenstate of $\hat{H}$, i.e.,  
\begin{equation}
\langle \cdot \rangle = \langle\Phi _0|\cdot|\Phi _0\rangle. 
\end{equation} 
In this case, we can write $\chi"(t)$ as  
\begin{eqnarray}
\chi "(t) &=& (2\hbar )^{-1}\left[ \langle\Phi _0|\hat{Q}(t)\hat{Q}(0) |\Phi_0
\rangle - \langle\Phi _0|\hat{Q}(0)\hat{Q}(t)|\Phi _0\rangle \right] \nonumber
\\ &=& (2\hbar )^{-1}\left[ e^{iE_0t/\hbar }\langle\Phi
_0|\hat{Q}e^{-i\hat{H}t/\hbar }\hat{Q}|\Phi _0\rangle
 - e^{-iE_0t/\hbar }\langle\Phi _0|\hat{Q}e^{i\hat{H}t/\hbar }\hat{Q}|\Phi_0
\rangle \right] \nonumber \\ &=& (2\hbar )^{-1}\sum_{n}\left[ e^{iE_0t/\hbar
}\ |\langle\Phi _0|\hat{Q}|\Phi _n\rangle|^2\ e^{-iE_nt/\hbar } - e^{-iE_0t/\hbar
}\ |\langle\Phi _0|\hat{Q}|\Phi _n\rangle|^2\ e^{iE_nt/\hbar }\right] 
\end{eqnarray}
where $\{|\Phi_n\rangle\}$ denote all the many-body eigenstates of the
isolated-system Hamiltonian,  
\begin{equation}
\hat{H}|\Phi _n\rangle = E_n|\Phi _n\rangle ,
\end{equation}
with $E_n \geq E_0$.

Upon Fourier transformation,
\begin{eqnarray}
{\tilde\chi}"(\omega ) &=& \pi \sum_n~|\langle\Phi _0|\hat{Q}|\Phi _n\rangle|^2
\ \left[ \delta(\hbar\omega +E_0-E_n)-\delta(\hbar\omega-E_0+E_n)\right]\nonumber
\\ &=& \sum_{n}~|\langle\Phi _0|\hat{Q}|\Phi _n\rangle|^2 ~\Im
\left(\frac{1}{\hbar \omega - E_0 + E_n + i0^{+}} - \frac{1}{\hbar \omega + E_0
- E_n + i0^{+}} \right) 
\end{eqnarray} 
where we used the identity
\begin{equation}
\frac{1}{x + i0^{+}} = {\cal P}\left( \frac{1}{x} \right) - i\ \pi\ \delta
(x) ,
\end{equation}
$\cal P$ denoting the Cauchy principal value. Methods have been developed to
evaluate semiclassically such expressions as above \cite{gab}. However, we
should recall that  the system is many-body so that such methods would require
the search for classical orbits of the many-body system. Simplification arises
by taking account of the Fermionic character of the system which allows a
reduction of the problem to one-body Hamiltonian in an effective potential
determined by, for instance, the Hartree-Fock method. We shall make this
simplifying assumption here and restrict our system to a set of uncoupled
one-body Hamiltonians. A similar assumption is carried out on the coupling
operator for which we assume the same form as for the Hamiltonian,
\begin{eqnarray}  
\hat{H} &=& \sum_{i=1}^{N}~\hat{h}_i, \nonumber \\  \hat{Q}
&=& \sum_{i=1}^{N}~\hat{q}_i 
\end{eqnarray} 
where $\hat{h}_i$ and $\hat{q}_i$ are one-body operators. Let us denote the
one-body eigenstates as  
\begin{equation} 
\hat{h}|\phi _a\rangle = \epsilon_a|\phi _a\rangle. 
\end{equation} 
When the Fermionic many-body system is in its
ground state, all the one-body eigenstates are occupied up to the Fermi energy,
$\epsilon _F$ so that the energy of the ground state is  
\begin{equation} 
E_0 = \sum_{a=1 \atop \epsilon _a \langle \epsilon _F}^{N} \epsilon _a.
\end{equation} 
In any many-body excited state, $|\Phi _n\rangle$, at least one
1-body level above the Fermi energy is occupied. We may denote such excited
states by the list of the occupied 1-body levels, or equivalently,  by the list
of the one-body levels for which the occupation is different with respect to the
ground state $|\Phi _0\rangle$, having in mind that such states are
antisymmetric for an exchange of two Fermions: 
\begin{eqnarray}
 |\Phi _0\rangle &=& |1111...11~{\vdots ^{\epsilon _{F}}}~00...0\rangle \nonumber
\\  |\Phi _n\rangle &=& |111...1110~{\vdots ^{\epsilon _{F}}}~0100...\rangle.
\end{eqnarray} 
For operators which are sums of 1-body operators as assumed in
(12),  we obtain the result that the matrix elements $\langle\Phi
_0|\hat{Q}|\Phi _n\rangle$ are non-vanishing only for the states $\Phi _n$ which
differ from the ground state $\Phi _0$ by one excitation. These states have a
hole in state $a$ ($\epsilon _a \langle \epsilon _F$) and a particle in state
$b$ ($\epsilon _b \rangle \epsilon _F$). For these states the matrix elements are
thus 
\begin{equation} 
\langle\Phi _0|\hat{Q}|\Phi _n\rangle = \langle\phi
_a|\hat{q}|\phi _b\rangle. 
\end{equation} 
Moreover the energy of the excited
state is 
\begin{equation} 
E_n = E_0 - \epsilon_a + \epsilon_b 
\end{equation}
under the assumption that $\epsilon _a \langle \epsilon _F \langle \epsilon_b$.
After some standard manipulations, we can write 
\begin{eqnarray}
&&\sum_{n}\frac{|\langle\Phi _0|\hat{Q}|\Phi _n\rangle|^2}{\hbar \omega - E_0 +
E_n + i0^{+}}  \nonumber \\ &=& \sum_{a,b}\frac{|\langle\phi _a|\hat{q}|\phi
_b\rangle|^2}{\hbar \omega - \epsilon _a + \epsilon _b + i0^{+}}\ [1 - \Theta
(\epsilon _F-\epsilon _b)]\ \Theta (\epsilon _F - \epsilon _a), 
\end{eqnarray}
where the Heaviside step function, $\Theta (\cdot)$ takes care of the
aforementioned restriction on the location of the 1-body states $\phi _a$ and
$\phi _b$ with respect to $\epsilon _F$.

In the presence of a thermal reservoir at temperature $T [=1/(\beta k_B)]$
described  with canonical density matrix, we get that   
\begin{eqnarray}
\langle\hat{Q}(t)\hat{Q}(0)\rangle &=& {1 \over Z}\ {\rm tr}~e^{-\beta
\hat{H}}e^{it\hat{H}/\hbar }\hat{Q}e^{-it\hat{H}/\hbar }\hat{Q} \nonumber \\ &=&
\sum_{mn} \exp [iE_m(t+i\beta \hbar )/\hbar ]\ \exp (-iE_nt/\hbar )\ 
|\langle \Phi_m|\hat{Q}|\Phi_n\rangle|^2. 
\end{eqnarray}
Thus the dynamical susceptibility is
\begin{equation}
{\tilde\chi}"(\omega ) = \frac{1}{Z(\beta )} \sum_{mn} 
|\langle\Phi_m|\hat{Q}|\Phi_n\rangle|^2 
\ \Im \frac{\exp(-\beta E_m)}{\hbar \omega - E_m + E_n + i0^{+}} - (\omega
\rightarrow -\omega ).  
\end{equation}  
Similar assumptions as before enable us to reduce this expression to the 1-body
system. Now the probability to find a state $|\Phi _m\rangle$ in which $|\phi
_a\rangle$ is occupied is given by the Fermi-Dirac distribution at energy
$\epsilon _a$. But $|\Phi _n\rangle$ is related to $|\Phi _m\rangle$ by the fact
that $|\phi _b\rangle$ must be unoccupied (see above). So, we have a joint
probability that $|\phi _a\rangle$ is occupied and $|\phi _b\rangle$ is
unoccupied, hence the propagator is  
\begin{equation} 
{\tilde\chi}"(\omega ) = \sum_{a,b} |\langle\phi _a|\hat{q}|\phi_b\rangle|^2 
\ \Im~\frac{ p^{FD}_a - p^{FD}_b}
{\hbar\omega - \epsilon _a + \epsilon _b + i0^{+}}  
\end{equation} 
where 
\begin{equation} 
p^{FD}_a = \frac{1}{\exp [\beta(\epsilon _a-\mu)] + 1} 
\end{equation} 
denotes the Fermi-Dirac probability or
mean occupation number. $\mu $ is the chemical potential fixed by the total
number of Fermions in the system. Eq. (21) gives therefore the Fourier transform
of the two-time correlation function.

At zero temperature, the Fermi-Dirac distribution becomes a Heaviside step
function, so the expression (21) becomes
\begin{equation}
{\tilde\chi}"(\omega ) = \sum_{a,b} |\langle\phi _a|\hat{q}|\phi
_b\rangle|^2\ \Im~\frac{\Theta (\epsilon _F-\epsilon _a) - \Theta (\epsilon
_F-\epsilon _b)}{\hbar \omega - \epsilon _a + \epsilon _b + i0^{+}}
\end{equation} 
which is identical with the expression previously derived. In the
foregoing discussion of this section, we have started from the time correlation
function (actually the commutator) and written an expression for the response.
Now that the susceptibility has been reduced to a 1-body expression, we can
treat the 1-body system semiclassically in terms of periodic orbits in the
effective Hartree-Fock Hamiltonian $\hat h$.

It is well to recall that ${\tilde\chi}"(\omega )$ has a physical interpretation
in terms of energy dissipation which is related to the fluctuations in the
frequency spectrum obtained as the Fourier transform of the time correlation
function. Indeed the fluctuation-dissipation theorem \cite{balescu} lies just in
realizing that connection. 

Employing the identity (11) in (21), and expanding the
difference between $p^{FD}$'s in the resulting expression, we obtain
\begin{eqnarray} 
{\tilde\chi}"(\omega ) &=& \pi \sum_{a,b} |\langle\phi
_a|\hat{q}|\phi _b\rangle|^2\ [p^{FD}(\epsilon _a) - p^{FD}(\epsilon _a + \hbar
\omega )]\ \delta (\hbar \omega + \epsilon _a - \epsilon _b)\nonumber \\ &=& -\pi
\sum_{n=1}^{\infty } \frac{(\hbar \omega )^{n}}{n!} \sum_{a,b}
\frac{\partial ^{n}p^{FD}}{\partial \epsilon ^{n}}(\epsilon _a)\ |\langle\phi
_a|\hat{q}|\phi _b\rangle|^2\ \delta (\hbar \omega + \epsilon _a - \epsilon _b)
\nonumber\\ &=& - \frac{1}{2\hbar} \sum_{n=1}^{\infty }\frac{(\hbar \omega
)^{n}}{n!}\ \tilde f_n(\omega).
\end{eqnarray}
where we introduced the expression
\begin{equation}
\tilde{f}_n (\omega ) = 2\pi\hbar \sum_{a,b}~\frac{\partial
^{n}p^{FD}}{\partial \epsilon ^{n}}(\epsilon _a)\ |\langle\phi _a|\hat{q}|\phi
_b\rangle|^2\ \delta (\hbar \omega + \epsilon _a - \epsilon _b)
 = \int dt\ e^{i\omega t}\ f_n(t) ,
\end{equation}
which is the Fourier transform of the time correlation functions
\begin{equation}
f_n(t) = {\rm tr}~\frac{\partial ^{n}p^{FD}}{\partial \epsilon
^{n}}(\hat{h}) \ e^{it\hat{h}/\hbar }\ \hat{q}\ e^{-it\hat{h}/\hbar }\ \hat{q}.
\end{equation}
Moreover, we can rewrite these time correlation functions in terms of a single
time correlation function as follows
\begin{equation}
f_n(t) = \int d\epsilon \frac{\partial ^{n}p^{FD}}{\partial \epsilon
^{n}}(\epsilon) \ C_{\epsilon}(t) , 
\end{equation}
with
\begin{equation}
C_{\epsilon}(t) ={\rm tr}\ \delta(\epsilon-\hat h) \ \hat q (t)\ \hat q(0) \qquad
\hbox{and} \qquad \hat q(t)  =  e^{it\hat{h}/\hbar }\ \hat{q}\
e^{-it\hat{h}/\hbar } . 
\end{equation}
In summary, we have so far reduced the many-body dynamical susceptibility to an
expression involving the time correlation function $C_{\epsilon}(t)$ of the
1-body effective dynamics:
\begin{equation}
{\tilde\chi}"(\omega) = -\pi \sum_{n=1}^{\infty} \frac{(\hbar \omega )^{n}}{n!}
\int \frac{dt d\epsilon}{2\pi\hbar}\ e^{i\omega t}\ \frac{\partial
^{n}p^{FD}}{\partial \epsilon ^{n}}(\epsilon) \ C_{\epsilon}(t) . 
\end{equation}
We will see in the next section that this expression is particularly suited
for developing semiclassical expansions.

\noindent{\bf 3.~Semiclassical Expressions of Response}

According to the preceding section, we need to write the semiclassical
expression for the time correlation function $C_{\epsilon}(t)$ in order to
obtain semiclassically the response function. To this end, we consider the
correlation function in the form:
\begin{equation} 
C_{\epsilon }(t) = {\rm tr}~\delta(\epsilon - \hat{h})\hat{X}
= {\rm tr}~\delta(\epsilon - \hat{h})\hat{A } + i\ {\rm tr}~\delta(\epsilon -
\hat{h})\hat{B}, 
\end{equation}
where $\hat X = \hat X(t)=\hat q(t) \hat q(0)$ is a 1-body operator.  This
operator is nonHermitian but can be decomposed as $\hat X= \hat A + i \hat
B$ in terms of two Hermitian operators: $\hat A = (\hat X +\hat X^{\dagger})/2$
and $\hat B= (\hat X -\hat X^{\dagger})/(2i)$.

Methods have been obtained to evaluate semiclassically such expressions in terms
of periodic orbits \cite{gab,efmw}. To motivate this method, we begin by
observing \cite{gr} that such an expression involves the matrix elements of
$\hat{A}$ (and $\hat{B}$) over the eigenstates of $\hat{h}$.  These matrix
elements can be obtained at the first-order perturbation theory of a perturbed
Hamiltonian $\hat h(\lambda)=\hat h+\lambda \hat A$. Assuming the eigenvalue
problem for $\hat{h}(\lambda )$ to be solved, the matrix elements of $\hat{A}$
may thus be obtained in terms of derivatives of eigenvalues of the perturbed
Hamiltonian, $\epsilon _n(\lambda )$, with respect to $\lambda $. Now the
periodic-orbit theory by Gutzwiller can be used to calculate the diagonal matrix
elements. 

Each term of the correlation function $C_{\epsilon }(t)$ can therefore be
expressed as 
\begin{eqnarray}
{\rm tr}~\delta (\epsilon -\hat{h})\hat{A} &=& -\frac{1}{\pi }\ \Im\ 
{\rm tr}~\frac{\hat{A}}{\epsilon -\hat{h} +i0^{+}} \nonumber \\  &=&
\frac{1}{\pi }\ \Im\ {\rm tr}~\frac{\partial }{\partial \lambda }\log (\epsilon -
\hat{h} - \lambda \hat{A} + i0^{+})\Big\vert_{\lambda = 0} . 
\end{eqnarray} 
Comparing with the following indentity
\begin{eqnarray}
-\frac{1}{\pi }\ \frac{\partial }{\partial \epsilon }\ \Im\ {\rm tr}~\log
(\epsilon - \hat{h} - \lambda \hat{A} + i0^{+})&=& {\rm tr}~\delta (\epsilon -
\hat{h} - \lambda \hat{A})\nonumber \\ &=& \frac{\partial }{\partial \epsilon }
N(\epsilon ; \lambda ), 
\end{eqnarray} 
which is the derivative of the staircase function $N(\epsilon\lambda)$ with
respect to the energy, we arrive at the relation, 
\begin{equation} 
{\rm tr}~\delta (\epsilon - \hat{h})\hat{A} = -\frac{\partial
N(\epsilon ; \lambda )}{\partial \lambda }\Big\vert_{\lambda = 0},
\end{equation}
which gives each term of the correlation function as the derivative of the
staircase function of the perturbed Hamiltonian $\hat h(\lambda)$ with respect
to $\lambda$.

On the other hand, one has the well-known semiclassical expression:
\begin{eqnarray}
N(\epsilon ;\lambda ) &=& \int \frac{d^fxd^fp}{(2\pi \hbar )^f}\ \Theta [\epsilon
- h_{W}(\lambda )] + O(\hbar ^{-f+1})\nonumber \\ &+& \sum_p\sum_{r=1}^{\infty }
\frac{1}{r\pi }\frac{\sin \left[ {r \over \hbar }S_p(\epsilon;\lambda )-r{\pi
\over 2}\nu _p\right]}{|\det[\hbox{\helv
m}_{p}^{r}(\lambda)-\hbox{\helv I}]|^{1/2}} + O(\hbar ), 
\end{eqnarray} 
where
$\hbox{\helv m}_{p}(\lambda)$ is the monodromy matrix governing the stability of
the classical periodic trajectory, $p$; $\nu_p$ is the Maslov index of the
trajectory and $h_{W}$ is the Weyl-Wigner transform  of the Hamiltonian 
$\hat{h}(\lambda )=\hat{h}+\lambda \hat{A}$. We notice that the above periodic
orbits $p$ are those of the perturbed Hamiltonian.  We assume here that
the periodic orbits of $h_W(\lambda)$ deform continuously to the
periodic orbits of $h_W(\lambda=0)$ for $\lambda$ small enough. For the
purpose of differentiating this expression with respect to $\lambda $, we
use the following classical formula which gives the derivative of the action of
the periodic orbits \cite{dd},
\begin{equation} 
\frac{\partial S_p}{\partial \lambda }=-\oint ~dt\ \frac
{\partial h_W(\lambda )}{\partial \lambda }=-\oint ~dt\ A_W, 
\end{equation} 
where the integrals go around the periodic orbit, $p$, and $A_W$ is the
Weyl-Wigner transform of the operator $\hat{A}$. Taking the derivative with
respect to $\lambda $ and adding both terms composing the correlation function,
we finally obtain   
\begin{eqnarray} 
{\rm tr}~\delta (\epsilon - \hat{h})\hat{X} &=& \int \frac{d^fxd^fp}{(2\pi \hbar
)^f}\ X_W\ \delta (\epsilon - h_{cl}) + O(\hbar ^{-f+1})\nonumber \\ &+&
\frac{1}{\pi \hbar }\sum_p\sum_{r=1}^{\infty}\frac{\cos \left({r \over \hbar
}S_p - r{\pi \over 2}\nu _p\right)}{|\det
(\hbox{\helv m}_p^r-\hbox{\helv I})|^{1/2}}\oint_{p}dt~X_W + O(\hbar ^{0})
\end{eqnarray} 
where $h_{cl}=h_W(\lambda =0)$ and the periodic orbits are those of $h_{cl}$.

For the case like ours, when
\begin{equation}
\hat{X} = \exp ({i \over \hbar }\hat{h}t)\ \hat{q}\ \exp (-{i \over \hbar
}\hat{h}t)\ \hat{q} \end{equation}
the Weyl-Wigner tranform can be written as 
\begin{eqnarray}
X_W({\bf x},{\bf p}) &=&\left( e^{{i \over \hbar }\hat{h}t} \right)_W e^{{i
\over 2}\hbar \hat\Lambda }\biggl\lbrace q_W e^{{i \over 2}\hbar \hat\Lambda }
\left[ \left( e^{{-i \over \hbar }\hat{h}t} \right)_W  e^{{i \over 2}\hbar
\hat\Lambda } q_W \right] \biggr\rbrace \nonumber \\ &=& \left[ \exp \left(
-\hat{{\cal L}}_{cl}t \right)q_W \right]q_W + ({\rm Weyl~~corrections}) \nonumber
\\ &=& q_W(t)q_W(0) + ({\rm Weyl~~corrections}) 
\end{eqnarray} 
where $\hat{{\cal L}}_{cl}=\{h,\cdot\}$ is the
classical evolution operator, and the operator $\hat\Lambda $ is standard
\cite{bjen} in the Weyl-Wigner expansions defined as  
\begin{equation} 
\hat\Lambda =
\frac{\overleftarrow\partial }{\partial {\bf p}}.\frac{\overrightarrow\partial
}{\partial {\bf x}} - \frac{\overleftarrow\partial }{\partial {\bf
x}}.\frac{\overrightarrow\partial }{\partial {\bf p}} ,
\end{equation}
and $q_W(t)=q_W\left[\Phi^t({\bf x},{\bf p})\right]$ is the value of $q_W$ at
the current point $\Phi^t({\bf x},{\bf p})$ of the trajectory from the initial
condition $({\bf x},{\bf p})$ of the Hamiltonian flow $\Phi^t$ of $h_{cl}$.

Putting all the pieces together, we can now write the semiclassical expansion
for the correlation function as  
\begin{eqnarray}
C_{\epsilon}(t) &=& {\rm tr}\ \delta (\epsilon - \hat{h})\ \hat{q}(t)\ \hat{q}(0)
\nonumber \\
&=&  \int \frac{d^fxd^fp}{(2\pi \hbar )^f}\ \delta\left[ \epsilon-h_{cl}({\bf
x},{\bf p})\right]\ q_W({\bf x},{\bf p})\ (e^{-\hat{{\cal L}}_{cl}t}q_W)({\bf
x},{\bf p})  + ({\rm Weyl~corrections}) \nonumber \\ &+& \frac{1}{\pi \hbar
}\sum_{p,r} \frac{\cos \left({r \over \hbar }S_p - r{\pi \over 2}\nu
_p\right)}{|\det (\hbox{\helv m}_p^r-\hbox{\helv I})|^{1/2}}\oint_{p}d\tau
~q_W(\tau )q_W(\tau + t) + O(\hbar ^0). 
\end{eqnarray} 
We now go back to Eq. (29) where we see that we still need to evaluate the
Fourier transform of the correlation function $C_{\epsilon}(t)$.

Let us evaluate the first term with $n=1$ of the susceptibility.  This should
give the dominant term of the Weyl series in powers of the Planck constant
$\hbar$ as well as a sum over periodic orbits with the smallest power in
$\hbar$ which gives therefore the most important contribution.  The other
periodic-orbit sums would involve smaller amplitudes with higher powers in
$\hbar$.  Replacing with the previous semiclassical result, we get
\begin{eqnarray}
{\tilde\chi}"(\omega) &=& - \frac{\omega}{2} \int d\epsilon\ \frac{\partial
p^{FD}}{\partial\epsilon}(\epsilon) \int dt\ e^{i\omega t}\ C_{\epsilon}(t) +
{\cal O}(\hbar) \nonumber\\
&=& - \frac{\omega}{2} \int dt \int \frac{d^fxd^fp}{(2\pi \hbar )^f}\ 
\partial_{\epsilon}p^{FD}\left[ h_{cl}({\bf x},{\bf p})\right]
\ q_W({\bf x},{\bf p})\ \left[e^{(i\omega -\hat{{\cal L}}_{cl})t}q_W\right]({\bf
x},{\bf p})\ +\ {\cal O}(\hbar^{-f+1}) \nonumber \\ &-& \frac{\omega}{2\pi\hbar}
\int d\epsilon\ \partial_{\epsilon}p^{FD}(\epsilon)
\sum_{p,r}  \frac{\cos \left({r \over \hbar }S_p - r{\pi \over 2}\nu
_p\right)}{|\det (\hbox{\helv m}_p^r-\hbox{\helv I})|^{1/2}} \int
dt\ e^{i\omega t}\ \oint_{p}d\tau ~q_W(\tau )q_W(\tau + t) + O(\hbar ^0) . 
\end{eqnarray} 
Assuming that $\hat q=\hat q^{\dagger}$ is a Hermitian operator, its
Weyl-Wigner transform is a real function of $({\bf x},{\bf p})$.  Along a
periodic orbit, we have the Fourier series
\begin{equation}
q_W(\tau) = \sum_{n=-\infty}^{+\infty} q_{p,n} \ \exp\left(i\frac{2\pi
n}{T_p}\tau \right) ,
\end{equation}
where $T_p$ is the classical period and $q_{p,-n}=q_{p,n}^*$.  We obtain that
\begin{equation}
\int dt\ e^{i\omega t}\ \oint_{p}d\tau ~q_W(\tau )\ q_W(\tau + t) = 2\pi\ 
T_p(\epsilon) \sum_{n=-\infty}^{+\infty} |q_{p,n}(\epsilon)|^2\ \delta\left[
\omega - \frac{2\pi n}{T_p(\epsilon)}\right] .
\end{equation}
Except in special systems like the harmonic oscillators, the periods
$T_p(\epsilon)$ vary with the energy $\epsilon$.  Because of the presence of
the Dirac distribution, we can perform the last integral over $\epsilon$ by
using
\begin{equation}
\delta\left[\omega - \frac{2\pi n}{T_p(\epsilon)} \right] = \frac{T_p^2}{2\pi
n\partial_{\epsilon}T_p}\ \delta\left[
\epsilon - \epsilon_{p,n}(\omega)\right] ,
\end{equation}
where $\epsilon=\epsilon_{p,n}(\omega)$ is the energy obtained by solving the
implicit equation $T_p(\epsilon)=2\pi n /\omega$.  We here assumed that the
periods vary monotonously with energy, otherwise we find a su of terms instead
of a single term in the right-hand member of the previous formula.  Noting that
\begin{equation}
\frac{dS_{p,n}}{d\omega} =
\frac{d}{d\omega}S_p\left[\epsilon_{p,n}(\omega)\right] = - \frac{ T_p^3}{2\pi
n\partial_{\epsilon}T_p} ,  \end{equation}
we have that
\begin{equation}
\int dt\ e^{i\omega t}\ \oint_{p}d\tau\  q_W(\tau )\ q_W(\tau + t) = 2\pi
 \sum_{n} |q_{p,n}(\epsilon)|^2
\ \Big\vert\frac{dS_{p,n}}{d\omega}\Big\vert\ \delta\left[ \epsilon -
\epsilon_{p,n}(\omega)\right] , 
\end{equation}
where $S_{p,n}=S_p\left[\epsilon_{p,n}(\omega)\right]$.

Substituting in the previous relation, we finally obtain the absorptive part of
the dynamical susceptibility as: 
\begin{eqnarray}
{\tilde\chi}"(\omega) &=& - \frac{\omega}{2} \int dt \int \frac{d^fxd^fp}{(2\pi
\hbar )^f}\ \partial_{\epsilon}p^{FD}\left[ h_{cl}({\bf
x},{\bf p})\right]\ q_W({\bf x},{\bf p})\ \left[ e^{(i\omega -\hat{{\cal
L}}_{cl})t}q_W\right]({\bf x},{\bf p})\ +\ {\cal O}(\hbar^{-f+1}) \nonumber \\
&-& \frac{\omega}{\hbar} \sum_{p,r,n}
 \partial_{\epsilon}p^{FD}\vert_{p,n} \ 
|q_{p,n}|^2\ \Big\vert\frac{dS_{p,n}}{d\omega}\Big\vert
\ \frac{\cos \left({r \over \hbar }S_{p,n} - r{\pi \over 2}\nu
_p\right)}{|\det (\hbox{\helv m}_{p,n}^r-\hbox{\helv I})|^{1/2}} \ +\ 
O(\hbar ^0) .  
\end{eqnarray} 
where all the expressions with indices ${p,n}$ are functions of the
frequency $\omega$ through their dependency on the energy
$\epsilon=\epsilon_{p,n}(\omega)$.  Eq. (47) is the central result of this
work.  

 To interpret this result, let us go back to the time correlation function
$C_{\epsilon}(t)$ which is
\begin{equation} 
{\rm tr}~\delta (\epsilon-\hat{h})\ \exp (i\hat{h}t/\hbar )\ \hat{q}\ \exp
(-i\hat{h}t/\hbar )\ \hat{q} = \frac{1}{2\pi}\int d\tau\ e^{i\epsilon\tau /\hbar}
\ {\rm tr}~e^{i\hat{h}(t-\tau )/\hbar}\ \hat{q}\ e^{-i\hat{h}t/\hbar}\ \hat{q}.
\end{equation}
In this expression, the trace can be rewritten in terms of the
Green propagators as  
\begin{eqnarray} &&{\rm tr}\ \exp [i\hat{h}(t-\tau )/\hbar
]\ \hat{q}\ \exp [-i\hat{h}t/\hbar ]\ \hat{q} \nonumber \\ &=& \int d{\bf x}d{\bf
x}'\ G({\bf x}',{\bf x};\tau -t)\ q({\bf x})\ G({\bf x},{\bf x}';t)\ q({\bf x}')
\nonumber \\ &=& \sum_{c,c'}\int d{\bf x}d{\bf x}'\ A_c({\bf x}',{\bf x};\tau
-t)\ \exp [iW_c({\bf x}',{\bf x};\tau -t)/\hbar ]\ q({\bf x})\nonumber \\
~~~~~~~~~~&\times& A_{c'}({\bf x},{\bf x}';t)\ \exp [iW_{c'}({\bf x},{\bf
x}';t)/\hbar ]\ q({\bf x}') 
\end{eqnarray} 
where in the last step, we have written
down symbolically the semiclassical expressions for the Green propagators. We
have here assumed that the operator $\hat q$ depends only on the position
operator, i.e., $\hat{q}=q(\hat{{\bf x}})$. The sum in the above
expressions is over classical paths, $c$ and $c'$. Notice that these are two
different paths in general as they arise from two different propagators. The
expression interprets for itself - from the point ${\bf x}$ to ${\bf x}'$, we
propagate along the path $c'$ for time $t$, and then we propagate back to the
point ${\bf x}$ along $c$ in time $\tau-t$. Precisely,  
\begin{eqnarray}
W_c({\bf x}'\rightarrow {\bf x};\tau -t) &=& \int_{0}^{\tau -t}~dt~L \nonumber
\\   W_{c'}({\bf x}\rightarrow {\bf x}';t) &=& \int_{0}^{t}~dt~L, 
\end{eqnarray}
$L$ being the Lagrangian of the system. The total phase in the correlation
function is the sum $W_c+W_{c'}$ of the two actions. By stationary phase method,
it can be shown that the two segments of the paths $c$ and $c'$, in fact, form a
periodic orbit of period $\tau $. This is the mechanism by which periodic orbit
corrections appear in the main result above. 

Classical contribution comes from the paths of zero length, i. e., $\tau = 0$,
and this is precisely the leading term in our expression, also called the
quasiclassical term. Thus, in any calculation of interest, we need to find the
classical correlation function, the Weyl corrections, and then the
periodic-orbit corrections. If the energy spectrum of the system is discrete, we
notice that the dynamical susceptibility (24) is a sum of Dirac peaks
centered on each of the Bohr frequencies $\omega = (E_m - E_n)/\hbar$. In the
semiclassical formula (47) the first term due to the paths of zero length
contribute to the dynamical susceptibility by a continuous function of frequency
$\omega $ which may be interpreted as the average or background of the discrete
quantum susceptibility. The second series of terms of (47) are due to the
periodic orbits which contribute by oscillatory functions of the frequency
$\omega$. Their sum is supposed to reproduce the Dirac peaks of the discrete
quantum susceptibility ${\tilde\chi}"(\omega )$.

We notice that, in contrast with the well-known Gutzwiller formula for the
level density, the quasiclassical term here involves a classical time
correlation function.  As a consequence, the spectral properties of the
classical Liouvillian $\hat{\cal L}_{cl}$ will intervene in the evaluation of
this term (as well as of the following Weyl corrections).  In this regard, we
mention that a periodic-orbit theory of the classical Liouvillian has been
developed \cite{ce,ga} in terms of classical zeta functions which resemble but
differ from the Selberg-Gutzwiller quatum zeta function.  Consequently, we have
here an example where the semiclassical evaluation of a quantum property,
namely ${\tilde\chi}"(\omega)$, involves the spectral properties of the
classical Liouvillian along with the Gutzwiller quantum periodic-orbit sum. 
It is remarkable that both these classical and semiclassical terms coexist within
the same formula, which points to the result that the classical periodic-orbit
theory is only one part of the underlying quantum dynamics.  We believe that
this result is general in the semiclassical evaluation of dynamical
susceptibilities, which include in particular the ac conductivity.

It turns out that the spectral properties of the Liouvillian differ
considerably whether the classical system $h_{cl}$ is integrable or chaotic.  

For integrable systems, the Liouvillian spectrum is formed by discrete
classical frequencies, $\hat{\cal L}_{cl} \psi_{\bf m}=i\Omega_{\bf m}\psi_{\bf
m}$, on each ergodic component which are the invariant tori.  The expression
would then be similar to the one obtained in \cite{dmb}.  The periodic-orbit
corrections should be given by the Berry-Tabor semiclassical theory under such
circumstances \cite{bt}.

For chaotic systems, the classical time correlation function in the
quasiclassical term decays if the system is mixing on the energy shell.  Recent
works have shown that this decay may be developed in terms of the complex
singularities of an analogue of the Liouvillian resolvent \cite{ce,ga,pr}.  When
the complex singularities are poles they are called Pollicott-Ruelle resonances
and are associated with exponential decays.  If the spectral decomposition of the
Liouvillian dynamics in terms of the Pollicott-Ruelle resonances is known the
quasiclassical term can be further reduced as a sum over the Pollicott-Ruelle
resonances over each energy shell \cite{ce,ga,pr}. 

\noindent
{\bf 4.~ Concluding Remarks }

 We have described a systematic procedure which lets us write important
quantities in many-body theory in a semiclassical expansion. All our
considerations are restricted to many-body systems reducible to  uncoupled
one-body systems, we believe that it is an interesting first step. We have shown
the central role played by the time correlation functions. On the one hand, they
are related by a Fourier transform to the response function; on the other hand,
they  facilitate useful expansions in terms of $\hbar $. The fact that the
Fermionic nature of the particles has been taken into account is an important
aspect. The other reason for an interest in the above considerations stems from
the fact that no assumptions are made about integrability of the underlying
classical one-body dynamics. It may be recalled that in a recent attempt along
similar lines \cite{dmb}, complete integrability was assumed. Moreover, in their
analysis, the Berry-Tabor formula for periodic orbit corrections has been used
which fails for the interesting case of the harmonic oscillators. Since we now
know that the Gutzwiller trace formula is exact for harmonic oscillators
\cite{bj}, and is also applicable for chaotic systems, we believe that our
analysis extends to  more general systems. 

There are many applications of such a formalism. In any discussion of
relaxation phenomena, time correlation functions play an important role. To
understand photoabsorption cross-section of atoms and molecules, semiclassical
treatment along these lines has been used \cite{gab,alonso}. In metallic
clusters and quantum dots, there is a growing interest in the semiclassical
treatment of the absorption spectra and cross-sections. The understanding of the
plasmon modes in different metallic clusters, their splitting etc. are believed
to have a semiclassical interpretation. However, for a systematic treatment of
the collective modes, we need to consider two-body interactions, thus a
Hamiltonian, 
\begin{equation} 
\hat{H} = \sum_{i} \hat{h}_i + \lambda \sum_{ij}\hat{v}_{ij} 
\end{equation} 
where $\hat{v}_{ij}$ denotes the two-body
term. What we need to do here is a perturbative treatment of the two-body
interaction by assuming that $\lambda \ll 1$ which, in turn, would give rise to
collective modes.

In a random-matrix framework, one can also study the correlation functions, and
it has been recently shown that the time correlations depend upon the
co-dimension of level crossing \cite{gjv}. A comprehensive understanding of all
the inter-relations will be one of the aims we look forward to. 

\newpage
\noindent
{\bf Acknowledgements}
S.R.J. is financially supported by the "Communaute Francaise de Belgique" under
contract no. ARC-93/98-166. S.R.J. is on an extraordinary leave from the
Theoretical Physics Division, Bhabha Atomic Research Centre, Bombay 400 085,
India. P. G. is grateful to the National Fund for Scientific Research
(F.~N.~R.~S. Belgium) for financial support.


\newpage 

\begin{thebibliography}{99} 

\bibitem{gab} P. Gaspard, D. Alonso,
and I. Burghardt, Adv. Chem. Phys. {\bf XC}, 105 (1995). 

\bibitem{bj} M. Brack
and S. R. Jain, Phys. Rev. A {\bf 51}, 3462 (1995). 

\bibitem{jl} S. R. Jain and S.
V. Lawande, Proc. Ind. Natl. Sc. Acad. {\bf 61} A, 275 (1995). 

\bibitem{gv} P. van Ede van der Pals and P. Gaspard, Phys. Rev. E {\bf 49}, 79
(1994). 

\bibitem{expt} J. A. Folk et al., Phys. Rev. Lett. {\bf 76}, 1699 (1996). 

\bibitem{brack} M. Brack, Rev. Mod. Phys. {\bf 65}, 677(1993). 

\bibitem{dmb} A. Dellafiore, F.
Matera, and D. M. Brink, Phys. Rev. A {\bf 51}, 914 (1995). 

\bibitem{jja} S. R. Jain, A. K. Jain, and Z. Ahmed, Phys. Lett. B {\bf 370}, 1
(1996). 

\bibitem{bdd} D. M. Brink, A. Dellafiore, and M. Di Toro, Nucl. Phys. A {\bf
456}, 205 (1986).

\bibitem{nhm} H. Nishioka, K. Hansen, and B. R. Mottelson, Phys. Rev. B {\bf 42},
9377 (1990). 

\bibitem{strut} V. Strutinsky, A. G. Magner, S. R. Ofengenden, and
T. Dossing, Z. Phys. A {\bf 283}, 269 (1977). 

\bibitem{mcg} M. C. Gutzwiller, {\em Chaos in Classical and Quantum Mechanics}
(Springer, New York, 1991).

\bibitem{balescu} R. Balescu, {\em Equilibrium and Nonequilibrium Statistical
Mechanics} (John Wiley and Sons, Inc., New York, 1974). 

\bibitem{fw} A. L. Fetter and J. D. Walecka, {\em Quantum Theory of Many-Particle
Systems} (McGraw Hill Inc., New York, 1971). 

\bibitem{efmw} B. Eckardt, S. Fishman, K. M\"uller, and D. Wintgen, Phys. Rev.
A {\bf 45}, 3531 (1992).

\bibitem{gr} This famous trick is first known to
have been applied by W. Pauli (Cf. Ref. [14]). The relevant reference for the
present context is the paper by P. Gaspard and S. A. Rice, Phys. Rev. A {\bf 48},
54 (1993).

\bibitem{ce} R. Artuso, E. Aurell, and P. Cvitanovi\'c, Nonlinearity {\bf 3},
325, 361 (1990); P. Cvitanovi\'c and B. Eckhardt, J. Phys. A: Math. Gen. {\bf
24}, L237 (1991).

\bibitem{ga} P. Gaspard and D. Alonso Ramirez, Phys. Rev. A {\bf 45}, 8383
(1992).

\bibitem{dd} M. L. Du and J. B. Delos, Phys. Rev. A {\bf 38}, 1896,
1913 (1988). 

\bibitem{bjen} N. L. Balasz and B. K. Jennings, Phys. Rep. {\bf
104}, 347 (1984). 

\bibitem{alonso} D. Alonso Ramirez, {\em Semiclassical
Quantization and Classically Chaotic Systems}, Ph. D. Thesis (Universit\'e Libre
de Bruxelles, unpublished, 1995).  

\bibitem{bt} M.V. Berry and M. Tabor, Proc. R. Soc. Lond A. {\bf 349}, 101-123,
(1976); M. V. Berry and M. Tabor, J. Phys. A: Math. Gen. {\bf 10}, 371
(1977).

\bibitem{pr} M. Pollicott, Invent. Math. {\bf 81}, 413 (1985); D. Ruelle,
Phys. Rev. Lett. {\bf 56}, 405 (1986); J. Stat. Phys. {\bf 44}, 281
(1986); J. Diff. Geom. {\bf 25}, 99, 117 (1987).

\bibitem{gjv} P. Gaspard, S. R. Jain, and P.
van Ede van der Pals, {\em Time Correlation Functions of Complex Quantum
Systems}~(preprint, 1996). 

\end{thebibliography}
\end{document}